\documentclass[amsmath,amssymb,aps,prl,superscriptaddress,final,twocolumn]{revtex4}
\usepackage{graphicx}
\usepackage{amssymb,amsmath,amsbsy}
\usepackage{epsf}

\usepackage{subfigure}

\usepackage{appendix}
\usepackage[sort&compress]{natbib}
\usepackage{color}

\newcommand{\tecpfigscale}{45mm}
\newcommand{\tecpfigsize}{40mm}
\citeindextrue
\begin{document}

\title{Nature-inspired microfluidic propulsion using magnetic actuation}

\author{S. N. Khaderi}

\affiliation{Zernike Institute for Advanced Materials, University of Groningen, Groningen, The Netherlands.}%

\author{M. G. H. M. Baltussen}
\author{P. D. Anderson}
\affiliation{Eindhoven University of Technology, Eindhoven, The Netherlands.}%

\author{D. Ioan}
\affiliation{Universitatea Politehnica din Bucuresti, Bucharest, Romania.}%

\author{J. M. J. den Toonder}
\affiliation{Eindhoven University of Technology, Eindhoven, The Netherlands.}%

\author{P. R. Onck\footnote{Corresponding author: p.r.onck@rug.nl}}
\affiliation{Zernike Institute for Advanced Materials, University of Groningen, Groningen, The Netherlands.}%

\begin{abstract}

In this work we mimic the efficient propulsion mechanism of natural cilia by magnetically actuating thin films in a cyclic but non-reciprocating manner.  By simultaneously solving the elasto-dynamic, magnetostatic and fluid mechanics equations, we show that the amount of fluid propelled is proportional to the area swept by the cilia. By using the intricate interplay between film magnetization and applied field we are able to generate a pronounced asymmetry and associated flow.  We delineate the functional response of the system in terms of three dimensionless parameters that capture the relative contribution of elastic, inertial, viscous and magnetic forces.

\end{abstract}
\maketitle

A rapidly growing field in biotechnology is the use of lab-on-a-chip devices to analyse bio-fluids \cite{0960-1317-14-6-R01, whitesides_nature, chang_nat_mat}. Such fluids have to be preprocessed (for example, mixed with other fluids \cite{den_toonder}) and transported to and from one or many micro-chambers where the biochemical analyses are performed. The microfluid transport through these stages is usually performed by downscaling conventional methods such as syringe pumps \cite{SchillingE.A._ac015640e, JeonN.L._la000600b},  micropumps \cite{0960-1317-14-6-R01}, or by exploiting electro-magnetic actuation, as in electro-osmotic \cite{shulin, lingxin} and magnetohydrodynamic devices  \cite{asuncion,  alexandra}. However, when transporting biological fluids (which usually have high conductivity), the use of electric fields may induce heating,   bubble formation and pH gradients from electrochemical reactions \cite{b408382m, wu:234103, henrik_2004_conference}. In this work, we explore a new way to manipulate  fluids in microfluidic systems, inspired by nature, through the magnetic actuation of artifical cilia.

Fluid dynamics  at the micrometer scale  is dominated by viscosity rather than inertia. This has important consequences for fluid propulsion mechanisms \cite{purcell}. In particular, mechanical actuation will only be effective in propelling fluids if their motion is cyclic, but asymmetric in shape change. Nature has solved this problem by means of hair-like structures, called cilia, whose beating pattern is asymmetric and consists of an effective and a recovery stroke \cite{murase}. While natural cilia use an internal forcing system based on motor proteins (dyneins), the key challenge for its artificial equivalent is the design of an externally-applied loading system that will generate a similar non-reciprocating motion. Recently, electrostatic artificial cilia have been experimentally shown to induce effective micro-mixing \cite{den_toonder}. In addition, magnetic fields are also used to induce flow,  but the asymmetry generated was found to be relatively small \cite{gauger_cilia}.
In this work we report on the identification of two simple magnetically-driven configurations that can create a large asymmetry. We will show that the fluid propelled is linearly proportional to the swept area by the film (the configurational space), which has been shown so far only for a non-actuated kinetic three-sphere model \cite{PhysRevE_69_062901}. The first is based on a magnetic instability that develops when the applied magnetic field is opposite to the direction of the magnetization in a permanently magnetic film. In a second configuration we will demonstrate that asymmetry can be achieved in a super-paramagnetic film, based on the intricate inter-play between the geometry of the film, the externally-applied field and the internally-induced magnetization.

The numerical model used in this work is based on a two-dimensional finite element representation of thin magnetic films, employing Euler-Bernoulli beam elements. We simultaneously solve for the elasto-dynamic equations of motion and Maxwell's equations, so that we can accurately account for the elastic, inertia and magnetic interaction in a non-linear geometry setting. We explicitly couple this Lagrangian solid-dynamics model to an Eulerian fluid dynamics model through Lagrange multipliers. Input to the fluid mechanics model are the positions and velocities of the film at all times which result in a full velocity field  in the fluid.  We calculate the drag forces on the film as tractions via the stress tensor in the fluid. The traction distribution is subsequently imposed as surface tractions in the magneto-mechanical model. A detailed description of the model can be found in the Supporting Information \cite{eom_support}. 

\begin{figure}[t]\centering
     \subfigure[]{\includegraphics[width=\tecpfigsize]{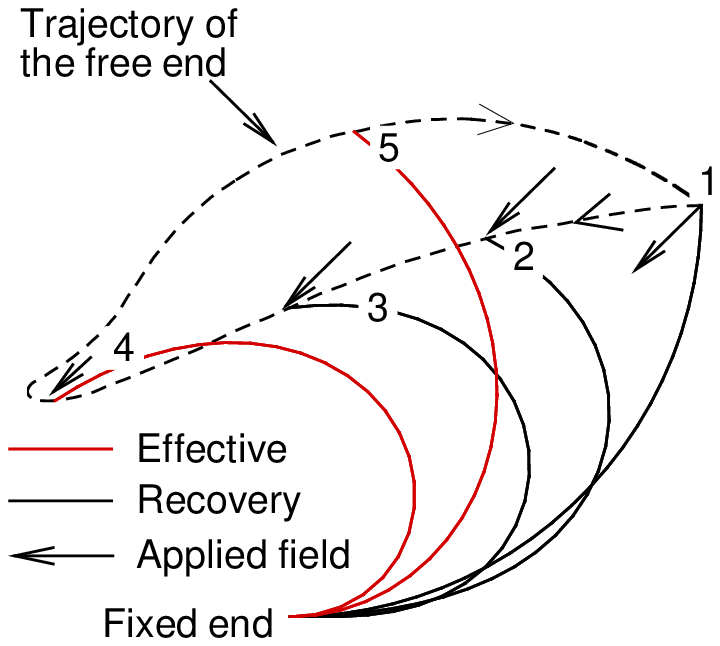}\label{fig:case2_summary_prl}}
     \subfigure[]{\includegraphics[width=\tecpfigsize]{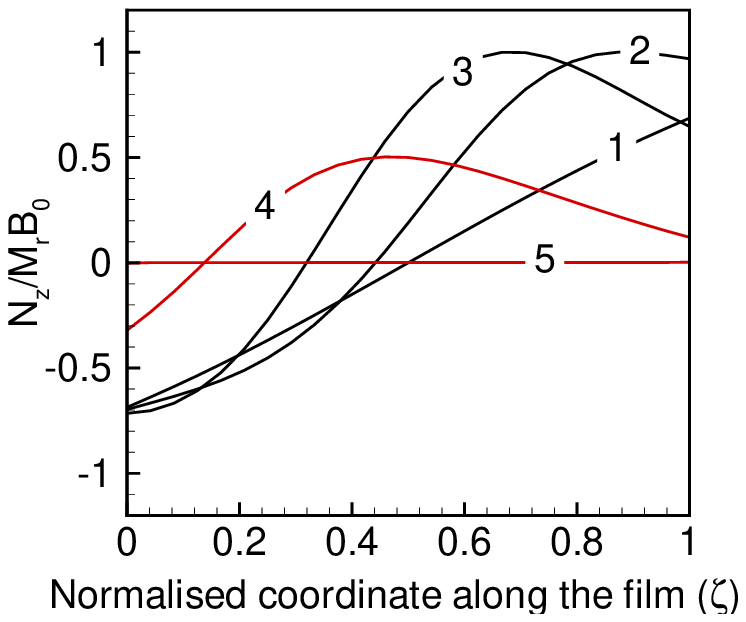}\label{fig:case2_torque_distribution}}
     \caption{Buckling of a curled permanently magnetic (PM) film as a result of magnetic actuation, during the propulsion of fluid. (a) Snapshots of the film at 
         0 ms, 0.3 ms, 0.6 ms, 1.1 ms and 3 ms. (b) Normalized torque distribution along the film corresponding to the snapshots shown in (a).}
     \label{fig:case2_total_figure_prl}
\end{figure}

We study a periodic arrangement of permanently magnetic (PM) cilia in a  microfluidic channel of height $5L$, with the cilia spaced $5L$ apart, where $L$ is the length of the cilia. A square unit-cell is identified consisting of one cilium. No-slip boundary conditions are applied at the top and bottom boundaries of the channel and periodic boundary conditions at the left and right ends of the unit-cell. The fluid has a viscosity $\mu=1\ \text{mPas}$. The film has a thickness $h=2$ $\mu$m, effective stiffness $E=\bar{E}/(1-\nu^2)=1$ MPa, where $\bar{E}$ is the elastic modulus and $\nu$ is the Poisson's ratio, and density $\rho=1600\ \text{kg/m}^3$. The initial geometry of the film is a quarter of a circle with radius $100\ \mu$m  fixed at the bottom of the channel, see instant 1 in Fig.~\ref{fig:case2_summary_prl}. The direction of the magnetization is along the film with the magnetization vector pointing from the fixed end to the free end. The remnant magnetization of the film is taken to be $M_r=15$ kA/m. A uniform external  field of magnitude $B_0=13.3$ mT is applied at $225^\circ$ to the $x$ axis from $t = 0$ ms to $t = 1$ ms and then linearly reduced to zero in the next $0.2$ ms. The results of the non-reciprocating motion of the film in the fluid during magnetic actuation are shown in Fig.~\ref{fig:case2_summary_prl}. The Eulerian fluid mesh is not shown for clarity. When the external field is applied, clockwise torques ($N_z$ is the magnetic body torque) are acting on the portion near the fixed end of the film while near the free end counter-clockwise torques develop (see instance 1 in Fig.~\ref{fig:case2_torque_distribution}). Under the influence of such a system of moments, the film undergoes a buckling kind of instability. This can be nicely seen from instances 1 and 2 in Figs.~\ref{fig:case2_summary_prl} and ~\ref{fig:case2_torque_distribution}. During this stage the position of zero torque is almost fixed, while the torques at the free end increase. This causes the film to snap through to configurations 3 and 4 during which the zero-torque position travels to the fixed end. Clearly, the initially opposing directions of the internal magnetization and the applied magnetic field are essential in generating an instability that causes a large bending deformation during application of the field.    Then, the applied field  is reduced to zero and the film returns to the initial position through instance 5 in Fig.~\ref{fig:case2_summary_prl}. Note that the propulsive action in the effective stroke (red) takes place during the elastic recovery of the film, while the film stays low in the recovery stroke due to the buckling-enforced snap-through.

\begin{figure}[t]\centering
     \subfigure[]{\includegraphics[width=\tecpfigsize]{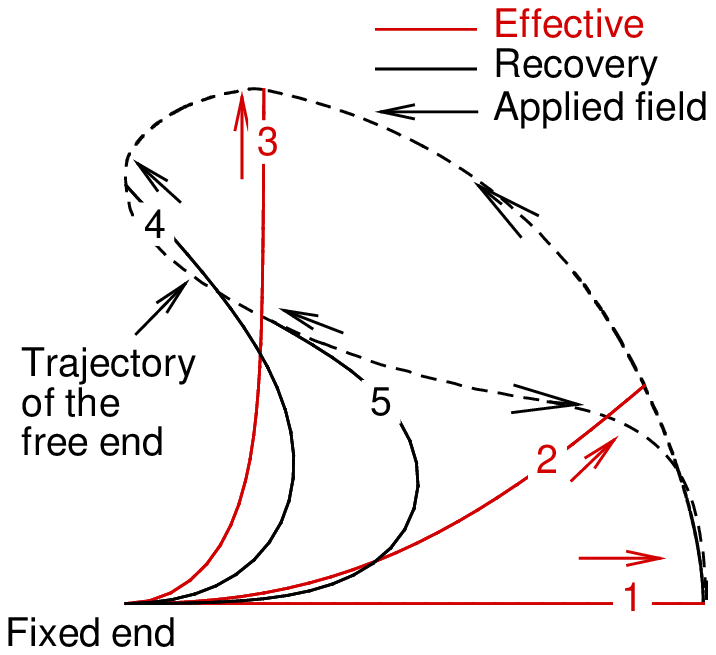}\label{fig:case3_summary_prl}}
     \subfigure[]{\includegraphics[width=\tecpfigsize]{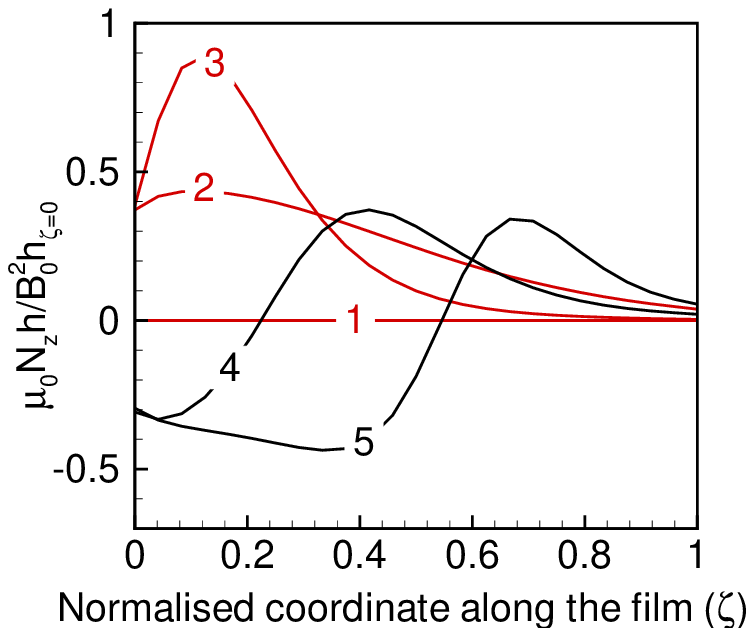}\label{fig:case3_torque_distribution}}
     \caption{Motion of a super-paramagnetic (SPM) film in a rotating magnetic field, during the propulsion of fluid. (a) Snapshots of the film at 
      0 ms, 2.5 ms, 5.0 ms, 7.5 ms and 8.5 ms. (b) Normalized torque distribution along the film corresponding to the snapshots shown in (a). Here $\mu_0$ is the permeability of vacuum, $h$ is the thickness at position $\zeta$ and $h_{\zeta=0}$ is the thickness at the fixed end.}
     \label{fig:case3_total_figure_prl}
\end{figure}

For a PM film the torques are maximum when the local magnetic film is perpendicular to the (remnant) magnetization. For a super-paramagnetic (SPM) film, however, the magnetization is induced by the field itself, posing different requirements on the applied magnetic fields in order to deform the film. A straight, magnetically anisotropic SPM film (having susceptibilities 4.6 and 0.8 in the tangential and normal directions, respectively), is subjected to a magnetic field with magnitude $B_0=31.5$ mT that is rotated from $0^\circ$ to $180^\circ$ in $t = 10$ ms and then kept constant during the rest of the cycle. The film has a length $L=100\ \mu$m, effective stiffness $E=1$ MPa and density $\rho=1600\ \text{kg/m}^3$.  Its cross-section is tapered, with the thickness varying linearly along its length, having $h=2\ \mu$m at the left (attached) end and $h=1\ \mu$m at the right end. Figure~\ref{fig:case3_summary_prl} shows that in the effective stroke the portion of the beam near the free end is nearly straight. This is due to the fact that in this region the film can easily follow the applied field so that field and magnetization are almost parallel, causing the magnetic torque to be  low in this region of the film (instances 2, 3 and 4 in Fig.~\ref{fig:case3_torque_distribution}).  When the film has reached position 4, the magnetization in the film is such that the torques are oriented clockwise near the fixed end and anticlockwise near the free end, resulting in strong bending of the film. From Fig.~\ref{fig:case3_torque_distribution} it can be seen that during the recovery stroke (in black) the position of zero torque propagates from the fixed end to the free end (from instance 4 to 5). Here the tapering is essential, causing the torque per unit length to be higher at the fixed end, allowing the film to recover to the initial position (1). This behaviour is very similar to that of natural cilia \cite{murase}. It is to be noted that the film recovers in the presence of an applied magnetic field. This sensitive interplay between stored elastic energy and controlled applied field can be exploited to provide a large asymmetry in motion.

\begin{figure}[h]\centering
\includegraphics[width=\tecpfigscale]{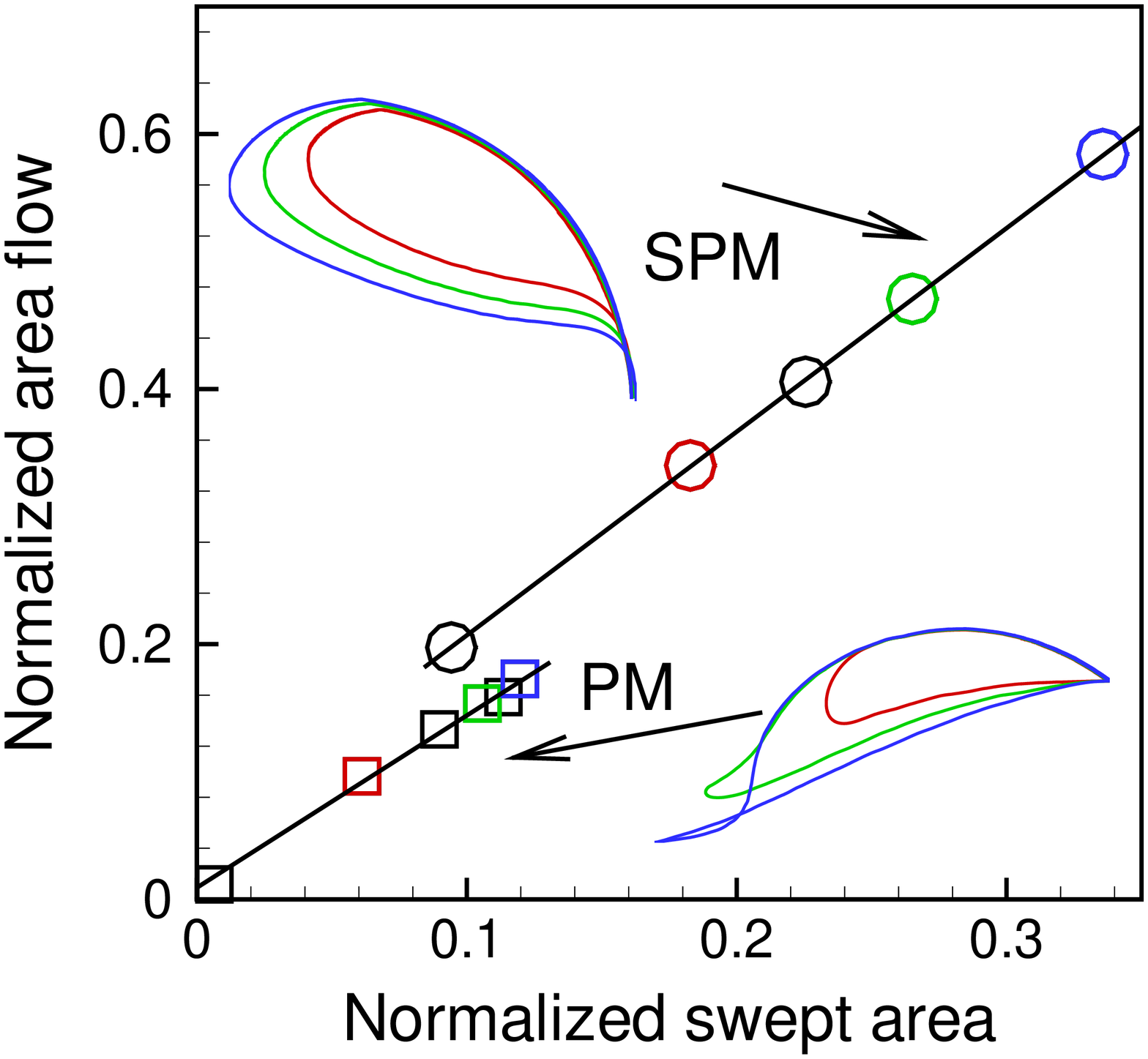}
     \caption{  Variation of normalized area flow with swept area. 
      }\label{fig:area_vs_flow}
\end{figure}

Next we analyze how much fluid is propelled by the two cases analyzed. We record the fluid volume transported through the channel per cycle and per unit out-of-plane thickness, giving an area flow per cycle. As a measure for the asymmetry, we compute the area swept by the free end of the film during one cycle (i.e. the area enclosed by the dashed lines in Figs. 1(a) and 2(a)) and vary this area by tuning the magnitude of the applied magnetic field (all other parameters remain unchanged).  Fig.~\ref{fig:area_vs_flow} shows the area flow per cycle as a function of the swept area for several different cases. We have normalized both quantities by the maximum area that the tip can sweep, $\pi L^2/2$. For three values of the magnetic field we plot the film tip trajectories for the PM and SPM configurations. The cycle times are 35 and 10 ms, respectively. The flux across the channel shows a linear dependence on the swept area. Similar result has been shown in \cite{golestanian:036308} where it is shown that the velocity of a three sphere swimmer is proportional to the area swept in the configurational space.
Due to the linear correlation between the swept area and the fluid flow, the swept area can be used as a measure of effectiveness of the actuator, representing the fluid volume displaced. This allows uncoupling the magneto-mechanical motion of the cilia from the computationally-intensive fluid dynamics calculations. Instead, we account for the fluid by means of velocity-proportional drag forces (using resistive force theory \cite{Johnson_brokaw}) on the cilia, with the drag coefficients calibrated to the coupled solid-fluid model (see \cite{eom_support}). 


To identify the dimensionless parameters that govern the behavior of the system, we start from the virtual work equation for the film \cite{eom_support}, neglecting the axial deformations:
$
      \int EI v''\delta v'' dx +
      \int\rho A\ddot{v}\delta v dx 
      -\int N_z \delta v'A dx
      +\int C_y\dot{v}\delta v b dx=0,
$
where $(\ )'={\partial (\ )}/{\partial x}$, $\dot{(\ )}={\partial (\ )}/{\partial t}$,   $I=bh^3/12$ is the second moment of area with $b$ the out-of-plane thickness, $A=bh$ is the cross-sectional area of the film and $v$ is the transverse displacement. In the virtual work equation 
 the first, second, third and last terms respectively represent the virtual work done by the elastic  internal bending moments, the inertial forces,  the magnetic couple and  the fluid drag forces. We introduce the dimensionless variables $V$, $T$ and $X$, such that $v=VL$, $x=XL$ and $t=Tt_\text{ref}$, where $L$ is a characteristic length (taken to be the length of the film) and $t_\text{ref}$ a characteristic time.
Substitution of these variables in the virtual work equation 
and normalization with the elastic term reveals the three governing dimensionless numbers: 
the inertia number, $I_n=12\rho L^4/Eh^2t_\text{ref}^2$, i.e. the ratio of inertial to elastic force, the magnetic number $M_n=12 N_z L^2/Eh^2 $, i.e. the ratio of magnetic to elastic force and the fluid number $F_n=12C_yL^4/Eh^3t_\text{ref}$, the ratio of fluid to elastic force. By substituting the torque expression for the two different magnetic materials,  the magnetic number $M_n$  for the PM film is linear in the applied field, $12M_rB_0L^2/Eh^2$, while  for the SPM film it is quadratic,  $12 B_0^2L^2/\mu_0Eh^2$.

\begin{figure*}[ht]
     \centering
     \subfigure[]{\includegraphics[width=\tecpfigscale]{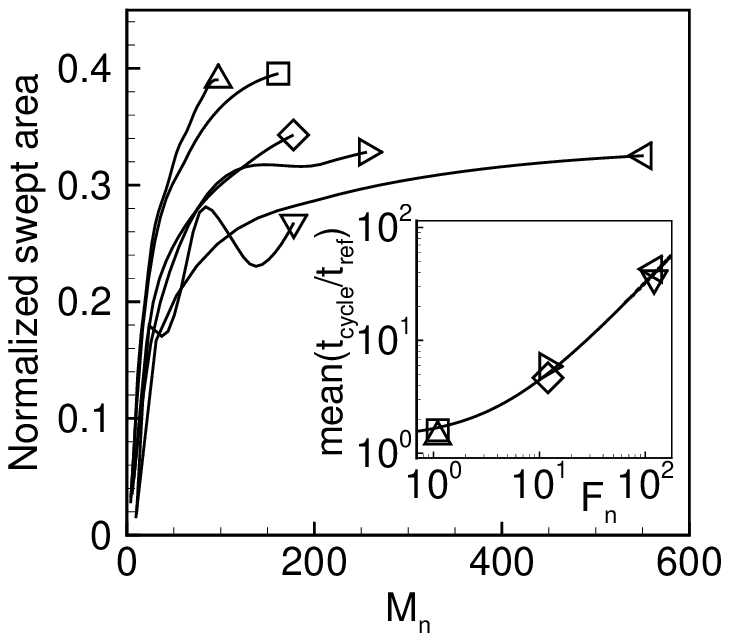}\label{fig:case2_area_vs_mn}}
     \subfigure[]{\includegraphics[width=\tecpfigscale]{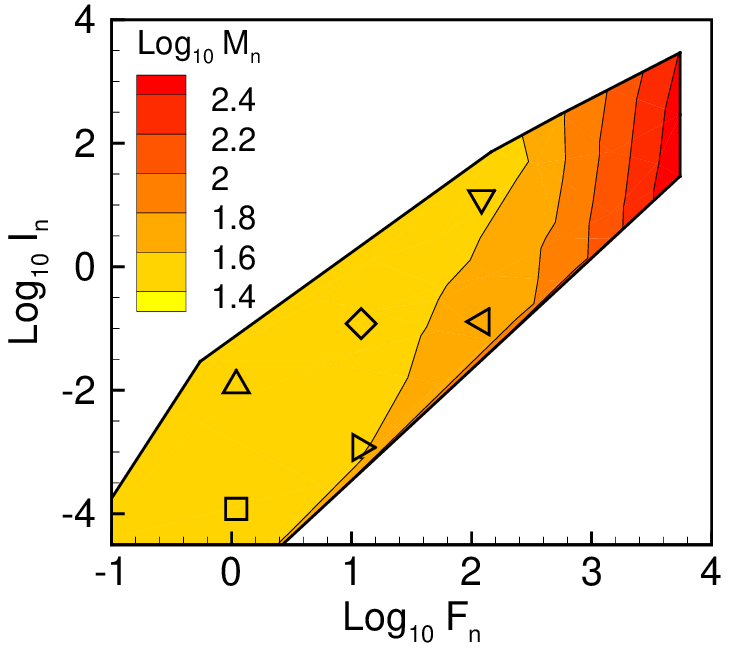}\label{fig:case2_mn_area_p2}}
     \subfigure[]{\includegraphics[width=\tecpfigscale]{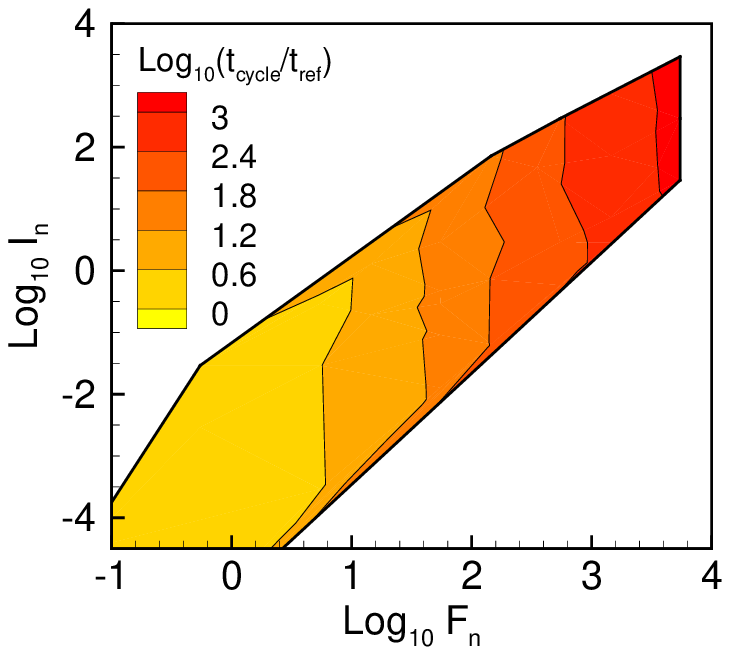}\label{fig:case2_mn_time_p2}}
     \subfigure[]{\includegraphics[width=\tecpfigscale]{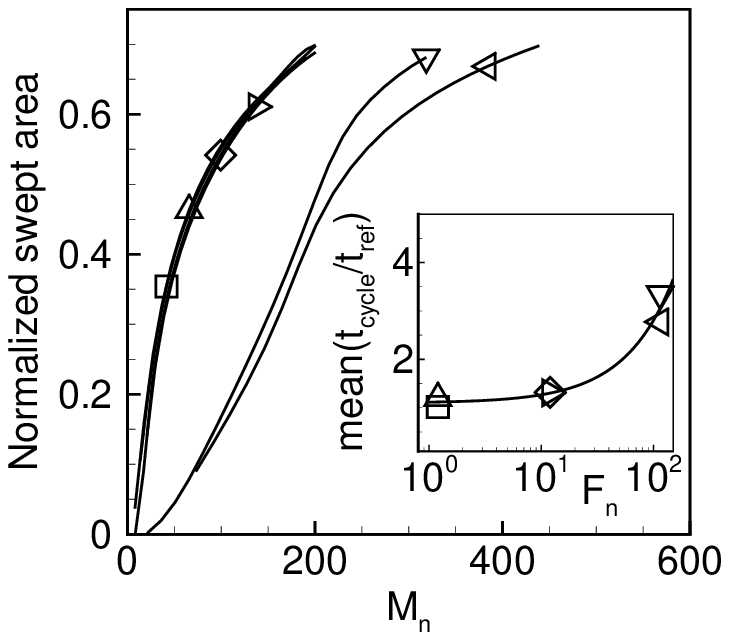}\label{fig:case3_area_vs_mn}}
     \subfigure[]{\includegraphics[width=\tecpfigscale]{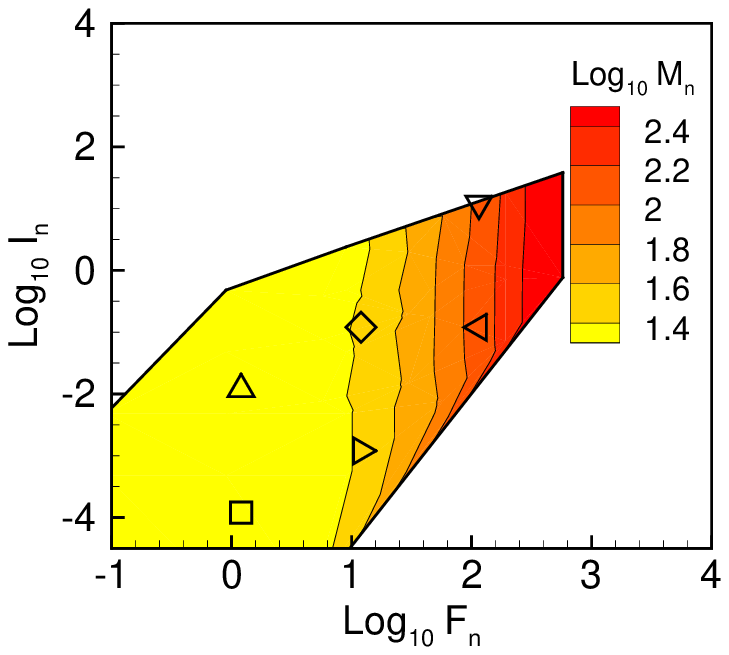}\label{fig:case3_mn_area_0p45}}
     \subfigure[]{\includegraphics[width=\tecpfigscale]{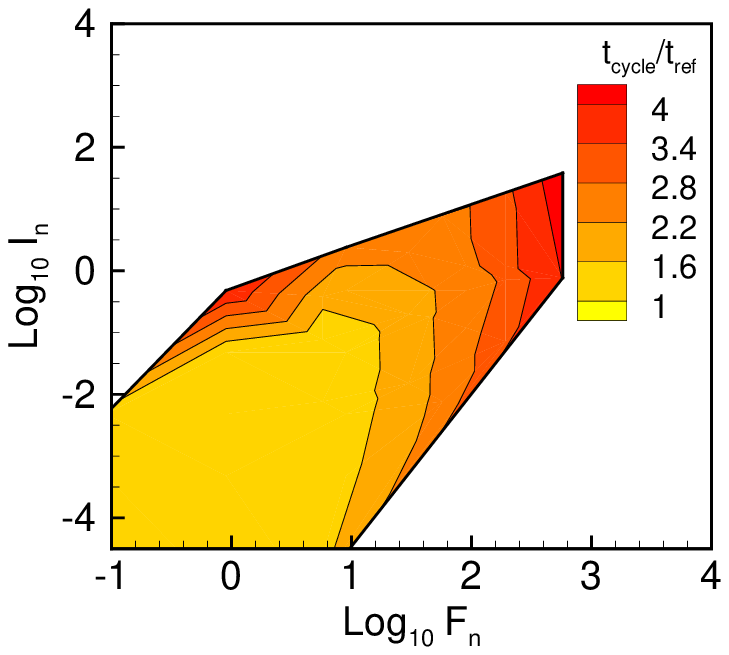}\label{fig:case3_time_area_0p45}}
\caption{Functional response of the PM system (top row, \subref{fig:case2_area_vs_mn}-\subref{fig:case2_mn_time_p2}) and the SPM system (bottom row, \subref{fig:case3_area_vs_mn}-\subref{fig:case3_time_area_0p45}). \subref{fig:case2_area_vs_mn} \subref{fig:case3_area_vs_mn} Normalized swept area as a function of $M_n$ for several combinations of $I_n$ and $F_n$, corresponding to the symbols of Figs. \subref{fig:case2_mn_area_p2} and \subref{fig:case3_mn_area_0p45}. The inset shows the mean normalized cycle time as a function of $F_n$. The mean is obtained by averaging all times corresponding to the data points that make up the specific $M_n$-swept area curve. \subref{fig:case2_mn_area_p2} \subref{fig:case3_mn_area_0p45} Contours of $M_n$ needed to sweep a normalized area of 0.2 for a wide range of $I_n$ and $F_n$ values. \subref{fig:case2_mn_time_p2} \subref{fig:case3_time_area_0p45} Contours of normalized cycle time corresponding to \subref{fig:case2_mn_area_p2} and \subref{fig:case3_mn_area_0p45}.}
  
     \label{fig:case2_contours}
\end{figure*}

We proceed by exploring the functional response of the system in terms of the swept area and cycle time, in dependence of the three dimensionless parameters. We analyzed  many  different combinations of $I_n, M_n$ and $F_n$, the results of which are summarized in Figs. \ref{fig:case2_area_vs_mn}-\ref{fig:case2_mn_time_p2} for the PM system and in Figs. \ref{fig:case3_area_vs_mn}-\ref{fig:case3_time_area_0p45} for the SPM system. Figs. \ref{fig:case2_area_vs_mn} and \ref{fig:case3_area_vs_mn} show the swept area as a function of $M_n$ for several combinations of $I_n$ and $F_n$.
The combinations are indicated by the different symbols, corresponding to specific locations in Figs. \ref{fig:case2_mn_area_p2} and \ref{fig:case3_mn_area_0p45}. The effect of all three parameters can also be nicely summarised by  analyzing what magnetic number and cycle time is needed to sweep a normalized area of 0.2, for a given range of $I_n$ and $F_n$ values (see Figs. \ref{fig:case2_mn_area_p2}, \ref{fig:case2_mn_time_p2}, \ref{fig:case3_mn_area_0p45}, \ref{fig:case3_time_area_0p45}).
  The swept area increases with $M_n$ reaching a maximum of 0.4 for the PM system (see Fig. \ref{fig:case2_area_vs_mn}), while values of 0.7 can be reached by the SPM system (see Fig. \ref{fig:case3_area_vs_mn}). For both systems the $M_n$ needed strongly increases with $F_n$. In other words, for a given elastic parameter set, larger magnetic forces are needed to overcome the drag forces imposed by the fluid (see Figs. \ref{fig:case2_area_vs_mn}, \ref{fig:case2_mn_area_p2}, \ref{fig:case3_area_vs_mn} and \ref{fig:case3_mn_area_0p45}).  It can be seen from Figs. \ref{fig:case2_area_vs_mn} and \ref{fig:case3_area_vs_mn} that the effect of $F_n$ is gradual for the PM system, while for the SPM case it is absent for small $F_n$, but suddenly kicks in for $F_n$ larger than 10. In addition, the inertial forces assist in generating asymmetry for both cases, although for the SPM system inertial effects are only triggered for very large fluid numbers (see Fig. \ref{fig:case3_area_vs_mn}). 

 For  the analysis of the normalised cycle time the reference time  was taken to be the time during which the field was applied (PM) or  rotated (SPM). It was observed that the cycle time dependence on $M_n$ and $I_n$ was very weak and mostly completely absent. The only clear dependence found was on $F_n$ which we show in the inset of Figs. \ref{fig:case2_area_vs_mn} and \ref{fig:case3_area_vs_mn} in terms of the normalized cycle times averaged over all different $M_n$ values analyzed and in Figs. \ref{fig:case2_mn_time_p2} and \ref{fig:case3_time_area_0p45} as the time required to sweep a normalised area of 0.2. 
 For the PM case the effective stroke is generated through the elastic recovery of the deformed film, without noticeable effect of inertial forces. For such overdamped systems the time taken by the system to return to the initial position scales linearly with $F_n$, the ratio of fluid to elastic forces.
The variation of normalized cycle time with $F_n$ for the SPM case is much smaller. This is due to the fact that the total cycle is performed in the presence of magnetic forces. For small $F_n$ and $I_n$ the mean normalized cycle time (see Fig. \ref{fig:case3_time_area_0p45}) is approximately equal to one; only for large $F_n$ and $I_n$ the cycle time is increased. At large $F_n$ the system relies on the recovery (going from instance 5 to instance 1 in Fig. \ref{fig:case3_total_figure_prl}) of the curved tip against high viscous forces (see Figs. \ref{fig:case3_area_vs_mn}, \ref{fig:case3_time_area_0p45}). The systems demonstrate an underdamped behaviour at large $I_n$ values causing inertial forces to generate large oscillations leading to a larger normalised time (see Fig. \ref{fig:case3_time_area_0p45}).  For a normalized area of 0.2 the response of both systems in the range $I_n < 0.1$ and $F_n < 10$ is quasi-static, i.e. independent of inertial and viscous effects.

To summarize, we have proposed and analysed  magnetic artificial cilia which can transport fluid in microfluidic channels. The main result is that we have found two simple and novel actuation mechanisms which can generate a pronounced  asymmetric motion of the cilia. One configuration is based on the buckling of a permanently magnetic film and the other is based on the intricate interaction between the applied field and the magnetization in a super-paramagnetic film. We have shown that the fluid propelled is linearly proportional to  the area swept by the film, which has so far only been shown for a non-actuated kinetic system \cite{PhysRevE_69_062901}. Finally, we have identified  the range of dimensionless parameters for which the artificial cilia exhibit an optimal behavior. The analysis presented  can be used as a guideline to make artificial cilia for microfluidic transport in lab-on-a-chip systems.

\begin{acknowledgments}
This work is a part of the $\text{6}^\text{th}$ Framework European project 'Artic', under contract STRP 033274.
\end{acknowledgments}

%

\end{document}